\documentclass[twocolumn,preprintnumbers,amsmath,amssymb]{revtex4}
\usepackage{graphicx}% Include figure files
\usepackage{dcolumn}% Align table columns on decimal point
\usepackage{bm}% bold math

\newcommand{\be}{\begin{equation}}
\newcommand{\bel}[1]{\begin{equation}\label{#1}}
\newcommand{\ee}{\end{equation}}
\newcommand{\bea}{\begin{eqnarray}}
\newcommand{\ba}{\begin{array}}
\newcommand{\eea}{\end{eqnarray}}
\newcommand{\ea}{\end{array}}

\begin{document}

\title{\bf Vehicular traffic flow at an intersection with the possibility of turning }

\author{M. Ebrahim Foulaadvand $^{1,2}$ \thanks{Corresponding Author: e-mail: foolad@iasbs.ac.ir,} and Somayyeh Belbasi $^{1}$}

\affiliation{ $^1$ Department of Physics, Zanjan University, P.O. Box 19839-313, Zanjan, Iran.}

\affiliation{$^2$ Computational Physical Sciences Laboratory, Department of Nano-Science, Institute for
Research in Fundamental Sciences (IPM), P.O. Box 19395-5531, Tehran,
Iran. }

\date{\today}
\begin{abstract}

We have developed a Nagel-Schreckenberg cellular automata model
for describing of vehicular traffic flow at a single
intersection. A set of traffic lights operating in fixed-time
scheme controls the traffic flow. Open boundary condition is
applied to the streets each of which conduct a uni-directional
flow. Streets are single-lane and cars can turn upon reaching to
the intersection with prescribed probabilities. Extensive Monte
Carlo simulations are carried out to find the model flow
characteristics. In particular, we investigate the flows
dependence on the signalisation parameters, turning probabilities
and input rates. It is shown that for each set of parameters,
there exist a plateau region inside which the total outflow from
the intersection remains almost constant. We also compute total
waiting time of vehicles per cycle behind red lights for various
control parameters.

\end{abstract}

\maketitle
\section{{Introduction}}

Simulation of urban traffic flow has shown to be of prime
importance for optimisation and control purposes as the number of
vehicles increases continuously and traffic conditions
deteriorate. Modelling traffic flow dynamics by cellular automata
has constituted the subject of intensive research by statistical
physics during the past years \cite{schadrev,helbingrev,tgf07}.
Understanding the characteristics of {\it city traffic} is one of
the most essential parts of the traffic research and was an early
simulation target for statistical physicists
\cite{bml,cuesta,nagatani1,freund,chopard,tadaki,torok,cs,brockfeld}.
Despite elementary attempts for simulation of perpendicular flows
\cite{nagatani3,ishibashi1,ishibashi2,foolad1,krbalek} the first
serious cellular automata describing the flow at a signalised
intersection was proposed by authors in \cite{foolad2}. Recently,
physicists have notably attempted to simulate traffic flow at
intersections and other traffic designations such as roundabouts
\cite{foolad3,foolad4,helbing2,helbing3,ray,chen,wang,huang,najem,foolad5,soh,bai1,bai2,du}.
Further recent progresses include various features such as
self-organised controlling of traffic lights and flows in
networks \cite{helbing08,helbing09}, the modelling of
decision-making at intersections \cite{fukui09}, mixture of
motorized and non motorized vehicles \cite{xie}, simulation of
traffic flow at T-shaped intersections \cite{li,ding}, data
collection in city traffic \cite{seo}, application of fuzzy logic
in signalisation of intersections \cite{nair}, visual simulation
of vehicular traffic \cite{hatzi} and dynamic route finding
strategy \cite{fang}. Vehicular flow at the intersection of two
roads can be controlled via two distinctive schemes. In the first
scheme, the traffic is controlled without traffic lights
\cite{foolad5,foolad6}. In the second, signalised traffic lights
control the flow. In Ref. \cite{foolad2}, we have modelled the
traffic flow at a single intersection with open boundary
conditions applied to the streets. In a recent study, a single
intersection operating under fixed time and traffic responsive
schemes has been explored with closed boundary condition
\cite{foolad7}. In the mentioned papers cars were restricted to
move straightly and not allowed to change their directions when
they reach to the intersection. In this work we incorporate the
possibility of turning. This incorporation constitutes a crucial
step for a more realistic description of traffic flow and would
certainly help us in a proper coordination of traffic lights.

\section{ Description of the Problem }

Consider the traffic flow at the intersection of two streets. Each
street conducts a uni directional flow and has a single lane.
Here for simplicity we exclude the possibility of overtaking
which requires having more than one lane and some laws for
lane-changing \cite{foolad8}. The flow directions are taken
south-north in street one and east-west in street two. Vehicles
can turn when reaching to the intersection. A northward moving
car can turn left and a westward moving car can turn right
(towards north) when reaching to the intersection. We model each
street by a chain of $L$ sites. These perpendicular chains
intersect each other at the middle sites $i_1=i_2=\frac{L}{2}$
(see Fig.1 for illustration). The discretisation of space is such
that each car occupies an integer number of cells denoted by
$N_{cell}$. Cell length is denoted by $\Delta x$. The car position
is denoted by the location of its head cell. Time elapses in
discrete steps of $\Delta t$ and velocities take discrete values
$0,1,2,\cdots, v_{max}$ in which $v_{max}$ is the maximum
velocity measured in unit of $\frac{\Delta x}{\Delta t}$. The
probability of turning a north-moving car to west is denoted by
$p_{sw}$. Correspondingly the probability of turning a west-moving
car to north is denoted by $p_{en}$.

\begin{figure}
\centering
\includegraphics[width=9.5cm]{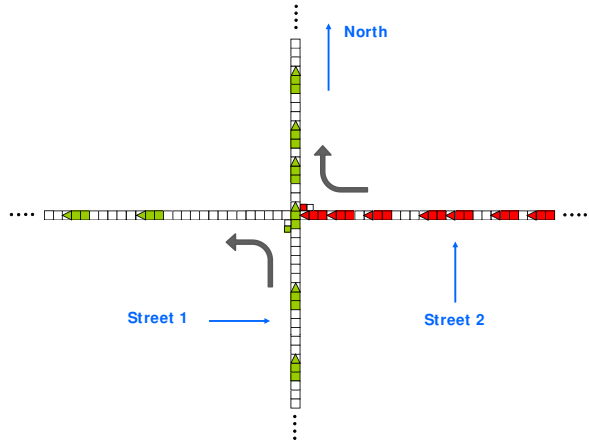}
\caption{ Intersection of two uni-directional single-lane streets with possibility of car turning at the intersection. } \label{fig:bz2}
\end{figure}

At each step of time, the system is characterized by the position
and velocity configurations of cars. The system evolves under the
Nagel-Schreckenberg (NS) dynamics \cite{ns}. Let us now specify
the physical values of our time and space units. The length of
each car, $L_{car}$, is taken 4.5 metres. Therefore, the spatial
grid $\Delta x$ (cell length) equals to $\frac{4.5}{N_{cell}}~m$.
We take the time step $\Delta t=1~s$. Furthermore, we adopt a
speed-limit of $74~ km/h$. In addition, each discrete increment
of velocity signifies a value of $\Delta v=\frac{4.5}{N_{cell}}
m/s $ which is also equivalent to the acceleration value. For
$N_{cell}=5$ we have $v_{max}=23$ cells per time step. Moreover,
the acceleration takes the comfort value $a=0.9 ~\frac{m}{s^2}$.
The value of random braking parameter is taken $p=0.1$ throughout
this paper. A set of traffic lights controls the traffic flow in
a fixed-time scheme as follows.

\subsection{Fixed Time Signalisation of lights}

In this scheme the traffic lights periodically turn into red and
green. The period $T$, hereafter referred to as {\it cycle time},
is divided into two phases. In the first phase with duration
$T_g$, the lights are green for the westward street and red for
the northward one. In the second phase which lasts for $T-T_g$
time steps the lights change their colour and become red for the
northward and green for the westward street. The gap of all cars
are updated with their leader vehicle except the ones (in each
street) which are the nearest approaching cars to the
intersection. These two cars need special attention. For these
cars gap should be adjusted with the signal in its red phase. In
this case, the gap is defined as the number of cells immediately
after the car's head to the intersection point $\frac{L}{2}$. If
the head position of the approaching car lies in the crossing
point the gap is zero. Note that for $N_{cell}>1$ at a time step
when a light goes red for a direction, portion of a passing car
from that direction can occupy the crossing point. In this case
the leading car of the queue in the other direction should wait
until the passing car from the crossing point completely passes
the intersection i.e.; its tail cell position become larger than
$\frac{L}{2}$.

\subsection{Modelling the car turning}

To model car turning we proceed as follows: once a car becomes approaching
to the intersection we draw a random number $r$ (uniformly chosen from the unit interval).
If $r$ is less than the corresponding turning probability the car turning label becomes one otherwise
it remains zero. In the subsequent time steps the approaching car's gap is adjusted appropriately:
if its turning label is zero its gap is adjusted via its forward leading car otherwise if its turning
label is one the gap is adjusted via the first car in the perpendicular direction which has already passed
the intersection (see figure 2).\\

\begin{figure}
\centering
\includegraphics[width=10.5cm]{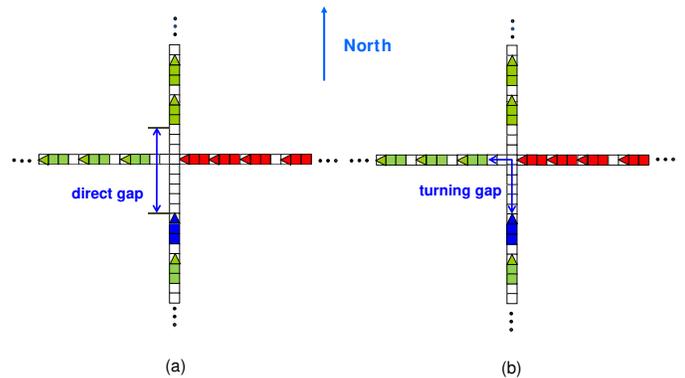}
\caption{ A turning adjusts its gap to the appropriate front car
in the destination lane. (a): the approaching car in street one
does not intend to turn. (b): the approaching car in street one
intends to turn. } \label{fig:bz2}
\end{figure}

\subsection{Entrance/Removal of cars to/from the intersection}

The type of boundary condition we implement in this paper is {\it
open}. the reason is that if one tries to imply close type
boundary condition, after some time steps most of the cars will
appear in that street which has a higher turning probability into
it. This effect arises because of accumulation of cars in the
street with higher inward turning probability. To avoid such
spurious effect which is surely contradicting to what occurs in
reality we impose open boundary condition. This type of boundary
condition is more compatible to realistic traffic flow. For this
purpose, we define two integer-valued parameters $d_1 \geq 1$ and
$d_2 \geq 1$ for streets one and two respectively. $d_i$ is
inversely proportional to the traffic density in street $i$. Once
the distance of the nearest car to first chain cell $j=1$
(farthest car to the intersection) in street $i~ (i=1,2)$ exceeds
$d_i$ a new car is inserted at site $j=1$ of street $i$ with
maximum velocity $v_{max}$. The removal of cars from a street is
modelled as follows: when the position of a car which has passed
the intersection becomes larger than $L-N_{cell}$ this car is
removed from the system and the number of exited cars is
increased by one correspondingly. The output current from street
$i$ is defined as the number of cars exited from street $i$
during a time interval $[T_1,T_2]$ divided by the $T_2 -T_1$. We
denote the output currents by $J_1$ and $J_2$
for street one and two respectively.\\

\section{Simulation results}

Now all the computational ingredients for simulation is at our disposal. We have taken the streets sizes equal to $1350~m$.
For $N_{cell}=1$ this corresponds to $L_1=L_2=300$~cells). The system is updated for $2\times10^5$ time steps. After transients,
two streets maintain steady-state currents $J_1$ and $J_2$ which are defined as the number of vehicles passing
the intersection per time step. They are functions of input rates $\frac{1}{d_1}$ and $\frac{1}{d_2}$, turning probabilities
$p_{sw}$, $p_{en}$ and signal times $T$ and $T_g$.

\subsection{ $N_{cell}=1$ }

In this case, each cell accommodates a car. Figure (3) shows
variation of $J_1$ versus green period of street two (east-west)
$T_g$ for some values of cycle time $T$. The maximum of $J_1$ is
less than $T_g=\frac{T}{2}$ because street two is less loaded
than street one. The inequality of input rates makes the slope
$J_1$ different at the vicinity of $\frac{T}{2}$. For a given set
of parameters, one can maximise the output current of street one
by setting $T_g$ to the maximising value. Sharp fall of $J_1$ is
noticeable and is related to enhancing of jammed region in street
one.

\begin{figure}
\centering
\includegraphics[width=7.5cm]{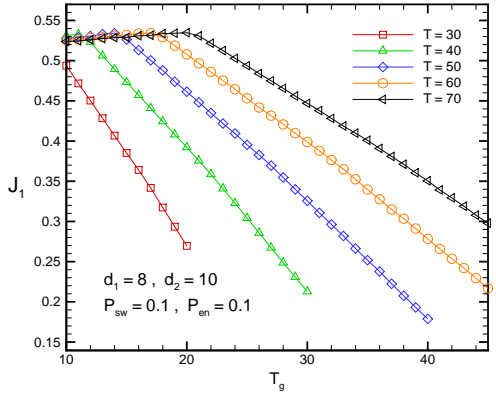}
\caption{ $J_1$ versus $T_g$ at $v_{max}=4$ for cycle time
$T=30,40,50,60$ and $70$ seconds. The values of other parameters
are specified in the figure. Street one has a higher input rate
than street two. Turning probabilities are equal to each other.}
\label{fig:bz2}
\end{figure}

Figure (4) depicts the behaviour of $J_1$ versus $T_g$ but this
time for various values of input parameter $d_1$. Other
parameters are specified in figure. By decreasing the green time
of street one, its output current slightly increases until it
reaches to plateau region on which $J_1$ is maintained constant.
The reason is that the red cycle of light optimally organises the
output flow from the intersection. This might be seem
counterintuitive to the expectation the less the red period the
larger the out flow. However, we should note that turning cars
from east-west street contribute to $J_1$ too.

\begin{figure}
\centering
\includegraphics[width=7.5cm]{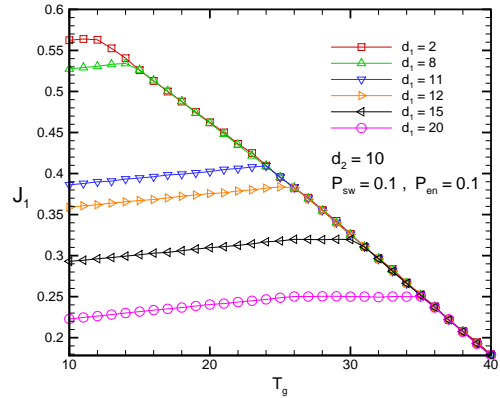}
\caption{ $J_1$ vs $T_g$ for various values of $d_1$ at $T=50$
and $v_{max}=4$. Other parameters values are specified in the
figure. } \label{fig:bz2}
\end{figure}

Figure (5) exhibits $J_2$ versus the the green time devoted to street two for various values of input rates of street one. Two distinctive
behaviour can be identified. For a given $d_1$ the current $J_2$ increases by increasing street two green time $T_g$ up to certain value.
Afterwards $J_2$ remain constant on a small plateau region and start to decrease in a slight manner afterwards. The reason is that by
increasing $T_g$ the green time of street one decreases. This reduces the current of south-west turning car into street two. In fact,
$J_2$ has two sources: straight moving east-west cars in street two and south-west turning cars in street one. Beyond a certain $T_g$ the
contribution of the second source dramatically falls which makes $J_2$ show a maximum.

\begin{figure}
\centering
\includegraphics[width=7.5cm]{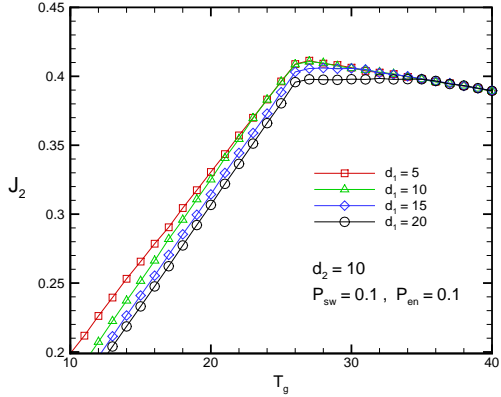}
\caption{ $J_2$ vs $T_g$ of street two for various values of $d_1$ at $v_{max}=4$. Other parameters values are specified in the figure. } \label{fig:bz2}
\end{figure}

Let us now look at the behaviour of the total flow $J_{tot}=J_1 +
J_2$ versus $T_g$ and $d_1$. Figure (6) shows this variation
versus street two green time $T_g$ for various values of cycle
time $T$. The notable point is that the maximum total flow only
weakly depends on the cycle time $T$. For each $T$ there is a
plateau region maintaining a maximum flow. This region is centered
at a $T_g=\alpha T$ and increases with cycle time $T$. $\alpha$
depends on the streets input rates and is less than $0.5$ when
$d_2$ is less than $d_1$. Roughly speaking $\alpha=\frac{d_1}{d_1
+ d_2}$. The slope is symmetrical on both side of the plateau
region.

\begin{figure}
\centering
\includegraphics[width=7.5cm]{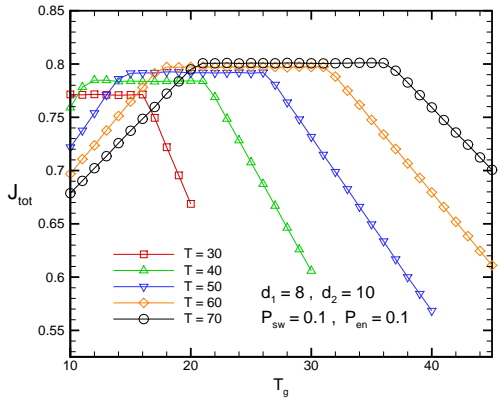}
\caption{ $J_{tot}$ vs $T_g$ of street two for various values of $T$ at $v_{max}=4$. Other parameters values are specified in the figure. } \label{fig:bz2}
\end{figure}

In figure (7) $J_{tot}$ is sketched versus $T_g$ this time for
various values of street one input rate $d_1$. Typically the
total flow shows a three-regime behaviour: an increasing, a
plateau and eventually a decreasing portion. The plateau length
decreases with increase in $d_1$. These three portions are
associated to the competition of two competing features. The
direct current and the turning current. The plateau region
corresponds to the situation where two features are balanced and
equally contribute to the total current.

\begin{figure}
\centering
\includegraphics[width=7.5cm]{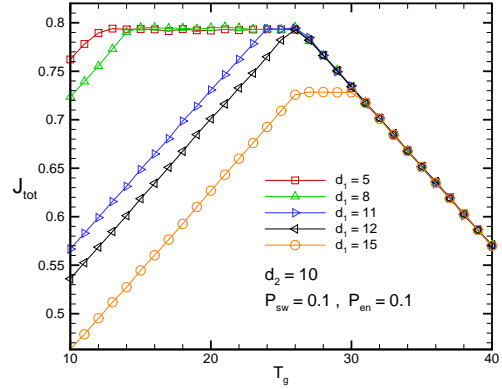}
\caption{ $J_{tot}$ vs $T_g$ for various values of $d_1$ at $v_{max}=4$. Other parameters values are specified in the figure. } \label{fig:bz2}
\end{figure}

To gain a deeper insight into the problem, we have obtained the
mutual dependence of the total flow $J_{tot}$ on input rate
parameters $d_1$ and $d_2$ for equal turning probability
$P_{sw}=P_{en}=0.1$ in figure (8) (see the journal printed
version). Other parameters are specified in the figure. We see the
existence of a two dimensional plateau on which the total flow is
maximised. The maximum throughput flow is slightly less than
$0.8$. Moreover, the total current $J_{tot}$ is symmetric respect
to the plane $d_1=d_2$ as it should be. The staircase like
topography of the current surface and the appearance of semi
plateau regions are noticeable. We see the dramatic fall of the
total current in the vicinity of the plane $d_1=d_2$. This marks
the inefficiency of the fixed-time scheme in organising the flow
via equal distribution of cycle times to equally loaded streets.
In figures (9) (see the journal printed version) and (10) we have
sketched the $J_{tot}$ versus $T$ and $T_g$ for given input
rates. As you can see there is a relative large plateau on which
the total current is maximal. The shape, size and orientation of
this plateau show strong dependence on input rates $d_1$ and
$d_2$. The maximal current value poorly depends on input rates.

\begin{figure}
\centering
\includegraphics[width=7.5cm]{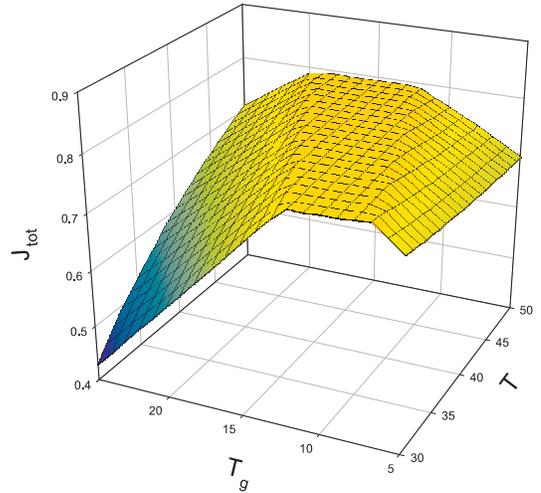}
\caption{ Three dimensional plot of $J_{tot}$ vs $T$ and $T_g$.
$d_1=7,d_2=15$, $v_{max}=4$ and turning probabilities are equally
set at $0.1$. $J_{tot}$ retains a constant value on a wide plateau
region and sharply falls outside the plateau.} \label{fig:bz2}
\end{figure}

Dependence of $J_{tot}$ on turning probabilities could serve us
to achieve a better understanding from the overall flow
characteristics of our single intersection. Figure (11) exhibits
such behaviour (see the journal printed version). The current
surface has interesting properties. There is a wide plateau
region maintaining a large current between 0.8 and 0.85. Sharp
decrease is followed when $|p_{sw}-p_{en}|$ becomes large. To see
the reason consider a situation where $p_{sw}$ is large whereas
$p_{sw}$ is small. Therefore most of the entrant cars to street
two move straight forward to west. On the other hand most of the
entrant cars to street one prefer to turn left (towards west)
when reaching to the intersection. This causes a dramatic jam in
the vicinity of the intersection which leads to a sharp decrease
of the total current. In figure (12) (see the journal printed
version) we have depicted a similar 3D plot this times for
unequal input rates $d_1 \neq d_2$. Other parameters are
identical to those in figure (11). The prominent effect of having
unequal input rates is the reduction of the maximal current value
in the plateau region from $0.82$ roughly to $0.71$.  Other
characteristics are analogous to the case $d_1=d_2$. Comparison
of figure (10) and (11) suggests that fixed-time signalisation
can poorly adjust itself to variation of turning probabilities
and implementation of more efficient adaptive strategies is
inevitable.

{\bf Waiting~ times}: \\

Despite maximising the output currents is highly desirable for us but it may not be the ultimate task in optimisation of traffic flow.
Another important parameter for optimisation of the traffic flow at intersections are aggregate waiting times of cars stopping at queues
 formed in the red cycle of the traffic lights. In our model this total waiting time, denoted by $WT_{tot}$ onwards, can be simply be evaluated.
 Once a car reaches to the end of a queue the $WT_{tot}$ counter is increased by one at each time step. This counting is paused once the
 corresponding light goes green. In figure (13) we show the $WT_{tot}$ per cycle versus green time of street two for various values of
 cycle time $T$. The system has evolved for $10^5$ timesteps. The first $20000$ timestep has been discarded for equilibration.
 Contrary to total current dependence on $T_g$, here we do not have a minimal plateau region but rather there exists a global minimum
 for $WT_{tot}$. This is helpful to us in achieving the optimisation of traffic light by simultaneous maximising the total current and
 minimisation of total waiting time. The results of figure (13) and (6) suggest to take the
 $T_g$at the end of current maximal plateau as the optimal choice for the green time distribution.

\begin{figure}
\centering
\includegraphics[width=7.5cm]{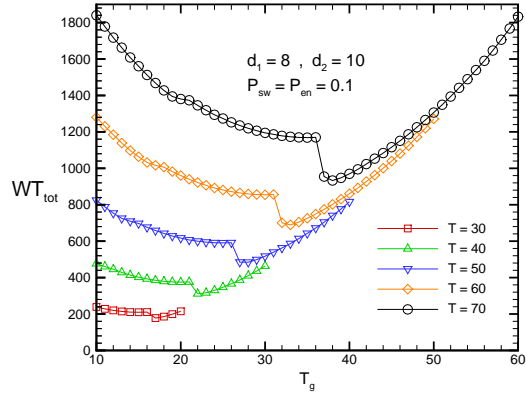}
\caption{ Total waiting time per cycle vs $T_g$. Street one
accommodates a heavier traffic flow ($d_1=8, d_2=10$). $v_{max}=4$
and turning probabilities are equally taken $0.1$. }
\label{fig:bz2}
\end{figure}

To gain a deeper insight, we have drawn a 3D plot of the
$WT_{tot}$ versus various control parameters in figures (14) and
(16). In figure (14) (see the journal printed version) the mutual
dependence of $WT_{tot}$ on input rates $d_1$ and $d_2$ for
$T=30$ and $T_g=15$ is plotted. The qualitative topography of the
$WT_{tot}$ surface resembles much with figure (8). For both $d_1$
and $d_2$ larger than $13$ (light traffic in both streets) we
have a relatively small $WT_{tot}$. Once either of $d_1$ or $d_2$
becomes less than $11$ a dramatic increase in $WT_{tot}$ is
observed. Figure (15) sheds more light onto the problem. This
figure is a 2D view of figure (14) and clearly shows the
staircase-like structure of $WT_{tot}$ surface. The sharp fall of
$WT_{tot}$ is noticeable. In figure (16) (see the journal printed
version) total waiting time per cycle is mutually sketched versus
turning probabilities for fixed equal input rates. We see there
is a wide minimal plateau on which the total waiting time retains
its constant minimal value. On the boundaries where
$|p_{sw}-p_{en}|$ becomes large the waiting time dramatically
begins to increase. The behaviours of total current and total
waiting times are complementary to each other i.e.; the maximum
of $J_{tot}$ coincides with the minimum of $WT_{tot}$. it is our
expectation that such complimentary between $WT_{tot}$ and
$J_{tot}$ remains valid for the cases in which the input rates
are not equal.

\begin{figure}
\centering
\includegraphics[width=7.5cm]{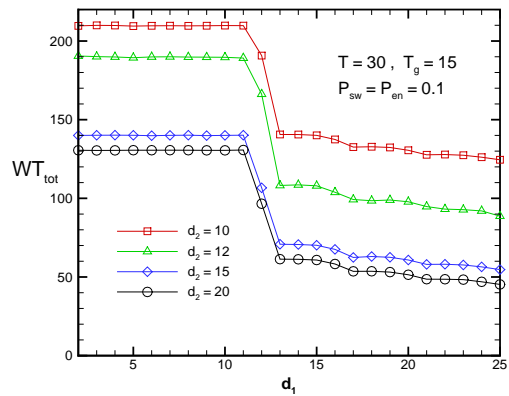}
\caption{ Dependence of $WT_{tot}$ per cycle on $d_1$ for fixed
values of $d_2$. Other parameters: $v_{max}=4$, $T=30$, $T_g=15$
and turning probabilities are equally taken $0.1$. }
\label{fig:bz2}
\end{figure}

We end this section by showing 3D plots of to $WT_{tot}$ mutually
versus $T$ and $T_g$. In figure (17) (see the journal printed
version) $WT_{tot}$ is given for $d_1=d_2=8$. There is no minimal
plateau sharply separated from high values. The curve is very
smooth and low value region gently crosses over to high values of
waiting times. When considering non equal input rates, the
situation becomes worse. In figure (18) (see the journal printed
version) the $WT_{tot}$ is sketched for $d_1=7\neq d_2=15$. Once
again we see the the region of maximal total current does not
entirely overlap with the region of minimal waiting time. This
demonstrates the optimisation of traffic flow is not a simple
task at least within the framework of NS model. Obviously
empirical data is needed and can illuminate the problem.

\subsection{ $N_{cell}=5$ }

In this section we show the results for a multi-cell occupation
of cars. Here we take $N_{cell}=5$ as discussed in section II. In
this case we have the comfort acceleration $a=0.9~ m/s^2$. We
have reproduced figures 3-7 to see the similarities/differences
imposed by varying $N_{cell}$. Figures (19-20) show exactly the
same graphs as in figures (3-4). The only difference is due to
$N_{cell}$ which is now equal to $5$.

\begin{figure}
\centering
\includegraphics[width=7.5cm]{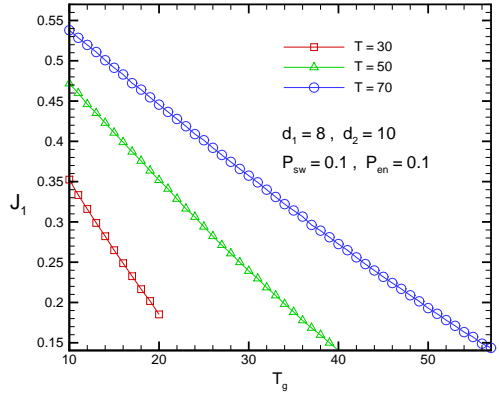}
\caption{ $J_1$ vs $T_g$ at $v_{max}=23$ for $N_{cell}=5$. The
values of other parameters are identical to figure (3). }
\label{fig:bz2}
\end{figure}

\begin{figure}
\centering
\includegraphics[width=7.5cm]{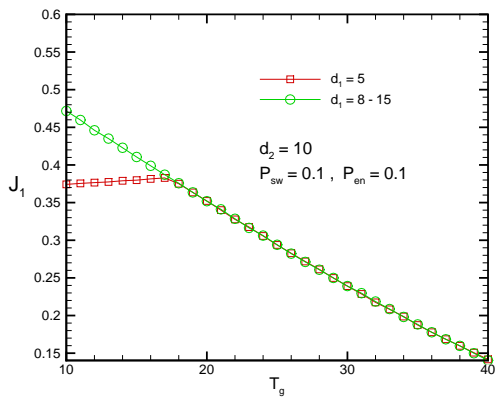}
\caption{ $J_1$ vs $T_g$ for various values of $d_1$ with
$N_{cell}=5$. Other parameters values are identical to figure
(19). } \label{fig:bz2}
\end{figure}

The most prominent difference is that in $N_{cell}=5$ the current
$J_1$ is decreasing whereas in $N_{cell}=1$ it slightly increases
with $T_g$ and then starts to decrease. Currents $J_1$ in uni-cell
i.e., $N_{cell}=1$ are in general larger than their counterparts
in multi-cell case. This is due to unrealistic large acceleration
in the NS model for uni-cell occupation. Figure (21) sketches the
same diagram shown in figure (5) for $N_{cell}=5$. As you see
$J_2$ is an increasing function of $T_g$ while in uni-call case
it reaches to a maximum, then after a short plateau it declines.

\begin{figure}
\centering
\includegraphics[width=7.5cm]{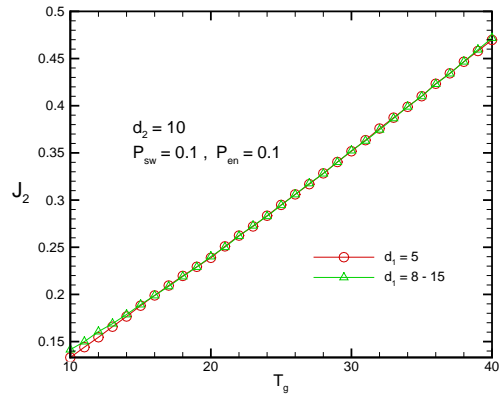}
\caption{ $J_2$ vs $T_g$ of street two for various values of
$d_1$ at $N_{cell}=5$. $v_{max}=23$. Other parameters values are
identical to figure (5). } \label{fig:bz2}
\end{figure}

Figures (22) and (23) exhibit dependence of total current $J_{tot}$ versus $T_g$ for various values of cycle time $T$ and input rate $d_1$.

\begin{figure}
\centering
\includegraphics[width=7.5cm]{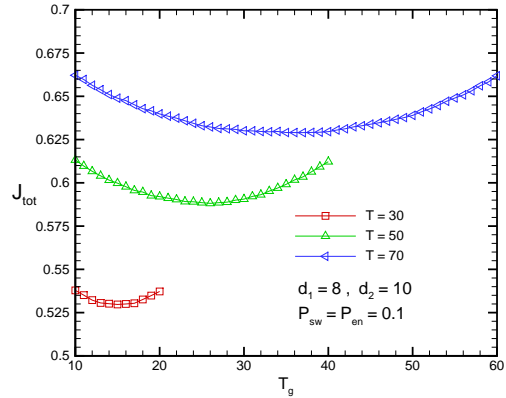}
\caption{ $J_{tot}$ vs $T_g$ of street two for various values of $T$ at $N_{cell}=5$. Other parameters values are identical to figure (6). } \label{fig:bz2}
\end{figure}

\begin{figure}
\centering
\includegraphics[width=7.5cm]{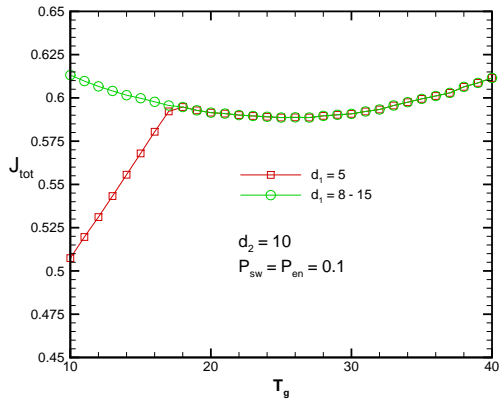}
\caption{ $J_{tot}$ vs $T_g$ for various values of $d_1$ at
$N_{cell}=5$. } \label{fig:bz2}
\end{figure}

Here also we encounter substantial differences. The most
distinguishable feature is the convex nature of total waiting
time curve. In sharp contrast to uni-cell case, in multi-cell
occupation, we do not have a plateau region. Instead a smooth
decrease is observed up to a minimum and then it symmetrically
begins to increase. Analogously, the total current value is less
than the uni-cell due to the same reason explained for $J_1$ and
$J_2$. In figure (23) and for $d_1=5$ The behaviour for small
$T_g$ is similar to uni-cell case i.e.; $J_{tot}$ increases. The
difference is that in uni-cell it reaches to a maximum and then
increase sharply whereas in multi-cell occupation $J_{tot}$ does
not decrease after linear increasing regime is ended. For $d_1>5$
we do not have the increasing behaviour for small $T_g$.

\section{Summary and concluding remarks}

In summary, we have simulated the vehicular traffic flow at a
single intersection in which the possibility of turning when cars
reach to the intersection is augmented to the problem. A set of
traffic lights operating in a fixed time scheme controls the flow.
We have tried to shed some light onto the problem of traffic
optimisation by extensive simulations. Total current is shown to
remain constant on a plateau of green time period given to one of
the directions. Three dimensional plot of total current for given
input rates of vehicles in terms of signalisation parameters
shows the existence of a 2D plateau region encompassed by almost
flat planes of sharp decreasing currents. By extensive
simulations we have examined the effect of turning on the output
current. The dependence of $J_{tot}$ on the whole range of
turning probabilities for fixed values of other parameters have
been computed including the equal and nonequal input rates. The
main finding is the appearance of a plateau current associated the
central region of the $P_{sw}-P_{en}$ plane. The current sharply
decreases when $|P_{sw}-P_{en}|$ becomes large. Needless to says
the plateau characteristics i.e.; its size and height depend on
the input rate and signalisation parameters $T$ and $T_g$.
Similar properties is observed if instead of $J_{tot}$ one looks
at total waiting time per cycle. Note here the waiting times
sharply increase when $|P_{sw}-P_{en}|$ becomes large. Besides
total current, total waiting time per cycle has been computed.
Our investigations reveal that in the parameter space, the
minimisation of total waiting time per cycle does not fully
coincide with the maximisation of total current. This arises the
natural question of {\it what quantity should be optimised in
order to acquire the most efficiency for the intersection ?} In
our Nagel-Schreckenberg cellular automata, not only the case of
uni-cell occupation, where each cell accommodates a car, has been
considered but also the more realistic case of multi cell
occupation in which a car occupies more than one cell has been
investigated. The corresponding results differ quantitatively and
in some case qualitatively. Seeking for advanced models properly
designated for modelling the behaviours of drivers at
intersections is inevitable and unavoidable. We would like to end
by emphasizing on the role of empirical data for adjusting the
parameters of any intersection traffic model. Despite utilising
multi-cell occupation can render the deceleration/acceleration
value of moving cars to a realist one, the behaviour of halted
cars in the queue when the light goes green is still suffering
from a satisfactory modelling.

\end{document}